\documentstyle[twoside,fleqn,espcrc2,amssymb,epsfig]{article}
\title{On the neutrino mass spectrum and neutrino mixing from oscillation data}
\author{
S.M. Bilenky\address{Joint Institute for Nuclear Research, Dubna, Russia, and
Institut f\"ur Theoretische Physik,
Technische Universit\"at Munchen,
D--85748 Garching, Germany},
C. Giunti\address{INFN, Sezione di Torino, 
and
Dipartimento di Fisica Teorica,
Via P. Giuria 1, I--10125 Torino, Italy}
and
W. Grimus\address{Institute for Theoretical Physics, University of Vienna,
Boltzmanngasse 5, A--1090 Vienna, Austria}
}
\begin{document}
\begin{abstract}
Two schemes of mixing of four massive neutrinos with two couples of close
neutrino masses separated by a gap of the order of 1 eV
can accommodate solar, atmospheric and LSND  neutrino oscillation data.            
It is shown that long-baseline $\bar\nu_e\to\bar\nu_e$
and $\nu_\mu\to\nu_e$
transitions are strongly suppressed in these schemes.
The scheme of mixing of three neutrino masses with a mass hierarchy that
can describe solar and atmospheric neutrino data is also discussed.
It is shown that in this scheme the effective Majorana mass
$|\langle{m}\rangle|$
that characterizes the matrix element of neutrinoless double-$\beta$ decay
is less than $ \sim 10^{-2} \, {\rm eV} $.\\
Preprint TUM-HEP-327/98, SFB 375-307, UWThPh-1998-50, DFTT 52/98, hep-ph/9809368.\\
Talk presented by S.M. Bilenky at \textit{Neutrino '98},
Takayama, Japan, June 1998.
\end{abstract}
\maketitle

\section{Introduction}
\label{sec1}

The conference \emph{Neutrino '98} is a very important event
in neutrino physics. At this conference
the Super-Kamiokande
collaboration~\cite{SK-atm}
presented the results of 535 days of measurement of
the atmospheric neutrino fluxes
which provide an impressive evidence for neutrino oscillations.

We discuss here which indications on
the neutrino mass spectrum
and on neutrino mixing
can be obtained from the results of Super-Kamiokande and all other
neutrino oscillation experiments. 
We will discuss also possible consequences for future experiments
that can be inferred from the analysis of the existing data.

In accordance with
the neutrino oscillation hypothesis (see~\cite{Bilenky})
the left-handed flavor neutrino fields
$\nu_{eL}$, $\nu_{{\mu}L}$ and $\nu_{{\tau}L}$
are linear combinations of the left-handed components of the
(Dirac or Majorana) massive neutrino
fields $\nu_i$:
\begin{equation}
\nu_{{\alpha}L}
=
\sum_{i} U_{{\alpha}k} \, \nu_{iL}
\,.
\label{01}
\end{equation}
In the LEP experiments on the
measurement of the invisible width of the $Z$-boson
it was proved that only three
flavor neutrinos exist in nature (see~\cite{PDG98}).
The number of massive neutrinos can be, however,
bigger than three (see~\cite{Bilenky}). 

If the total lepton number
$L=L_e+L_\mu+L_\tau$
is conserved,
the neutrinos $\nu_i$
are Dirac particles
and
the number of massive neutrinos is equal to three.  

If the total lepton number is not conserved,
the neutrinos $\nu_i$ are massive
Majorana particles
($\nu_i=\nu_i^c\equiv{\cal C}\overline{\nu_i}^T$,
where ${\cal C}$
is the charge-conjugation matrix).
In the general case,
the number of Majorana fields $\nu_i$
is $n=3+m$,
where $m$ is the number of right-handed
fields $\nu_{aR}$
that enter in the neutrino mass term.
We have in this case

\begin{equation}
\nu_{{\alpha}L}
=
\sum_{i=1}^{n}
U_{{\alpha}i} \, \nu_{iL}
\,,
\quad
(\nu_{aR})^c
=
\sum_{i=1}^{n}
U_{ai} \, \nu_{iL}
\,,
\label{02}
\end{equation} 
where $U$ is a $n{\times}n$ unitary mixing matrix.

Two possible options are usually discussed in the Majorana case:

\begin{enumerate}

\item
The \emph{see-saw} option~\cite{see-saw}.
If the total lepton number is violated by the right-handed Majorana mass term
at an energy scale much larger than the electroweak scale,
the Majorana mass spectrum is composed of
three light masses $m_i$ ($i=1,2,3$)
and three very heavy masses $M_i$ ($i=1,2,3$)
that characterize the scale of lepton number violation.
The light neutrino masses are given by the see-saw formula
\begin{equation}
m_i \sim \frac{ ( m_i^F )^2 }{ M_i }
\ll
m_i^F
\quad
(i=1,2,3)
\,.
\label{011}
\end{equation}
where $m_i^F$ is the mass of
the charged lepton or up-quark in the $i^{{\rm th}}$ generation.
The see-saw mechanism provides a
plausible explanation for the smallness of neutrino masses
with respect to the masses of all other fundamental fermions.

\item
The \emph{sterile neutrino} option.
If more than three Majorana mass terms are small,
then there are light sterile neutrinos.
In this case
active neutrinos
$\nu_e$, $\nu_\mu$ and $\nu_\tau$
can transform into sterile states $\nu_a$
that are quanta of right-handed fields
$\nu_{aR}$.
Notice that sterile neutrinos can be obtained in the framework of the
see-saw
mechanism with some additional assumptions
(``singular see-saw''~\cite{CKL98}).

\end{enumerate}

From the analysis of the data of atmospheric neutrino experiments it
follows that
$ \Delta{m}^2_{{\rm atm}} \sim 2 \times 10^{-3} \, {\rm eV}^2 $~\cite{SK-atm},
where
$\Delta{m}^2$
is the difference between the squares of neutrino masses.
Another scale of
$\Delta{m}^2$
was obtained
\cite{HL97,FLM97}
from the analysis of the data of all solar neutrino experiments:
$
\Delta{m}^2_{{\rm sun}}
\sim
10^{-5} \, {\rm eV}^2
$
(MSW~\cite{MSW})
or
$
\Delta{m}^2_{{\rm sun}}
\sim
10^{-10} \, {\rm eV}^2
$
(vacuum oscillations).
Finally,
indications in favor of
$\bar\nu_\mu\to\bar\nu_e$
and
$\nu_\mu\to\nu_e$
oscillations
with a third scale of $\Delta{m}^2$,
$
\Delta{m}^2_{{\rm LSND}}
\sim
1 \, {\rm eV}^2
$,
were obtained
in the accelerator LSND experiment~\cite{LSND}.

All these indications in favor of neutrino masses and mixing will
be checked by future solar, long-baseline and short-baseline neutrino
oscillation 
experiments (see these Proceedings).

We will consider two possible scenarios:

\begin{enumerate}

\item
All three indications in favor of neutrino
oscillations are confirmed.

\item
Only the indications of solar and atmospheric neutrino experiments
are confirmed.

\end{enumerate}

\begin{figure}[t!]
\begin{center}
\setlength{\unitlength}{1cm}
\begin{tabular*}{\linewidth}{cccc}
\begin{picture}(1.4,4)
\thicklines
\put(0.5,0.2){\vector(0,1){3.8}}
\put(0.4,0.2){\line(1,0){0.2}}
\put(0.8,0.15){\makebox(0,0)[l]{$m_1$}}
\put(0.4,0.4){\line(1,0){0.2}}
\put(0.8,0.45){\makebox(0,0)[l]{$m_2$}}
\put(0.4,0.8){\line(1,0){0.2}}
\put(0.8,0.8){\makebox(0,0)[l]{$m_3$}}
\put(0.4,3.5){\line(1,0){0.2}}
\put(0.8,3.5){\makebox(0,0)[l]{$m_4$}}
\end{picture}
&
\begin{picture}(1.4,4)
\thicklines
\put(0.5,0.2){\vector(0,1){3.8}}
\put(0.4,0.2){\line(1,0){0.2}}
\put(0.8,0.2){\makebox(0,0)[l]{$m_1$}}
\put(0.4,2.9){\line(1,0){0.2}}
\put(0.8,2.9){\makebox(0,0)[l]{$m_2$}}
\put(0.4,3.3){\line(1,0){0.2}}
\put(0.8,3.25){\makebox(0,0)[l]{$m_3$}}
\put(0.4,3.5){\line(1,0){0.2}}
\put(0.8,3.55){\makebox(0,0)[l]{$m_4$}}
\end{picture}
&
\begin{picture}(1.4,4)
\thicklines
\put(0.5,0.2){\vector(0,1){3.8}}
\put(0.4,0.2){\line(1,0){0.2}}
\put(0.8,0.2){\makebox(0,0)[l]{$m_1$}}
\put(0.4,0.6){\line(1,0){0.2}}
\put(0.8,0.6){\makebox(0,0)[l]{$m_2$}}
\put(0.4,3.3){\line(1,0){0.2}}
\put(0.8,3.25){\makebox(0,0)[l]{$m_3$}}
\put(0.4,3.5){\line(1,0){0.2}}
\put(0.8,3.55){\makebox(0,0)[l]{$m_4$}}
\end{picture}
&
\begin{picture}(1.4,4)
\thicklines
\put(0.5,0.2){\vector(0,1){3.8}}
\put(0.4,0.2){\line(1,0){0.2}}
\put(0.8,0.15){\makebox(0,0)[l]{$m_1$}}
\put(0.4,0.4){\line(1,0){0.2}}
\put(0.8,0.45){\makebox(0,0)[l]{$m_2$}}
\put(0.4,3.1){\line(1,0){0.2}}
\put(0.8,3.1){\makebox(0,0)[l]{$m_3$}}
\put(0.4,3.5){\line(1,0){0.2}}
\put(0.8,3.5){\makebox(0,0)[l]{$m_4$}}
\end{picture}
\\
(I) & (II) & (IIIA) & (IIIB)
\end{tabular*}
\end{center}
\begin{center}
Figure \ref{fig1}
\end{center}
\null \vspace{-1.5cm} \null
\refstepcounter{figure}
\label{fig1}
\end{figure}

\section{Four massive neutrinos}
\label{sec2}

At least four massive neutrinos are needed in order
to have three different scales of
$\Delta{m}^2$
\cite{four,BGKP,BGG96,BGG97a,BGG97-98,CKL98}.
The three types of neutrino mass spectra that can accommodate 
the solar, atmospheric and LSND scales of $\Delta{m}^2$
are shown in Fig.~\ref{fig1}.
In all these spectra
there are two groups of close masses
separated by a gap of the order of 1 eV
which gives
$
\Delta{m}^2_{41} \equiv m_4^2 - m_1^2
\simeq \Delta{m}^2_{{\rm LSND}}
$.

Only the
largest mass-squared difference
$ \Delta{m}^2_{41} $
is relevant
for the oscillations in short-baseline (SBL) experiments
and the SBL transition probabilities have the same
dependence on the parameter
$ \Delta{m}^2_{41} L / 2 p $
( $L$ is
the source-detector distance
and $p$ is the neutrino momentum)
as the standard
two-neutrino probabilities~\cite{BGKP}:
\begin{eqnarray}
&&
P_{\nu_\alpha\to\nu_\beta}
=
\frac{1}{2} \, A_{\alpha;\beta} \,
\left( 1 - \cos \frac{ \Delta{m}^2_{41} L }{ 2 p } \right)
\,,
\label{Ptran}
\\
&&
P_{\nu_\alpha\to\nu_\alpha}
=
1
-
\frac{1}{2} \, B_{\alpha;\alpha} \,
\left( 1 - \cos \frac{ \Delta{m}^2_{41} L }{ 2 p } \right)
\,.
\label{Psurv}
\end{eqnarray}
The oscillation amplitudes
$A_{\alpha;\beta}$ and $B_{\alpha;\alpha}$
depend on the elements on the mixing matrix $U$
and on the form of the neutrino mass spectrum:
\begin{eqnarray}
&&
A_{\alpha;\beta}
=
4
\left|
\sum_i
U_{{\beta}i}
\,
U_{{\alpha}i}^*
\right|^2
\,,
\label{Aab}
\\
&&
B_{\alpha;\alpha}
=
4
\left( \sum_i |U_{{\alpha}i}|^2 \right)
\left( 1 - \sum_i |U_{{\alpha}i}|^2 \right)
\,,
\label{Baa}
\end{eqnarray}
where the index $i$ runs over the indices of the first or
(because of the unitarity of $U$) second group
of neutrino masses.

The results of SBL reactor
$\bar\nu_e$
and
accelerator
$\nu_\mu$
disappearance
experiments in which no oscillations were found
imply that
$ B_{\alpha;\alpha} \leq B_{\alpha;\alpha}^0 $
for
$\alpha=e,\mu$.
The upper bounds
$B_{\alpha;\alpha}^0$
are given by the exclusion curves obtained from the data of 
SBL disappearance experiments
and depend on the value of $\Delta{m}^2_{41}$.
Using Eq.(\ref{Baa}),
these upper bounds
imply
the following constraints for
the quantities
$ \sum_i |U_{{\alpha}i}|^2 $
($\alpha=e,\mu$):
\begin{equation}
\sum_i |U_{{\alpha}i}|^2 \leq a_\alpha^0
\quad \mbox{or} \quad
\sum_i |U_{{\alpha}i}|^2 \geq 1 - a_\alpha^0
\,,
\label{05}
\end{equation}
where
\begin{equation}
a_{\alpha}^0
=
\frac{1}{2}
\left( 1 - \sqrt{ 1 - B_{\alpha;\alpha}^0 } \,\right)
\,.
\label{06}
\end{equation}
The most stringent values of
$a_e^0$ and $a_\mu^0$
can be obtained from the results of
the Bugey reactor experiment~\cite{Bugey95}
and of the CDHS and CCFR
accelerator experiments~\cite{CDHS-CCFR}.

We have considered
the range
$
10^{-1} \leq
\Delta{m}^2_{41}
\leq 10^3 \, {\rm eV}^2
$.
In this range
$ a_e^0 \lesssim 4 \times10^{-2} $
and
$ a_\mu^0 \lesssim 2 \times 10^{-1} $
for $ \Delta{m}^2_{41} \gtrsim 0.3 \, {\rm eV}^2 $
(see Fig.~1 of Ref.~\cite{BBGK96}).
Thus, from the results of disappearance experiments
it follows that
$ \sum_i |U_{ei}|^2 $
and
$ \sum_i |U_{{\mu}i}|^2 $
can be either small or large (close to one).

From the four possible sets of values of the quantities
$ \sum_i |U_{ei}|^2 $
and
$ \sum_i |U_{{\mu}i}|^2 $
(small-small, small-large, large-small and large-large),
for each neutrino mass spectrum only one set
is compatible with the results
of solar and atmospheric neutrino
experiments~\cite{BGKP,BGG96}.
In the case of spectra I and II we have
\begin{equation}
|U_{ek}|^2 \leq a_e^0
\quad \mbox{and} \quad
|U_{{\mu}k}|^2 \leq a_\mu^0
\,,
\label{07}
\end{equation}
with $k=4$ for spectra I and $k=1$ for spectra II.
In the case of spectrum IIIA we have
\begin{equation}
\sum_{i=1,2} |U_{ei}|^2 \leq a_e^0
\ \mbox{and} \
\sum_{i=1,2} |U_{{\mu}i}|^2 \geq 1 - a_\mu^0
\,,
\label{08}
\end{equation}
whereas in the case of spectrum IIIB we have
\begin{equation}
\sum_{i=3,4} |U_{ei}|^2 \leq a_e^0
\ \mbox{and} \
\sum_{i=3,4} |U_{{\mu}i}|^2 \geq 1 - a_\mu^0
\,.
\label{09}
\end{equation}

\begin{figure}[t!]
\begin{center}
\mbox{\epsfig{file=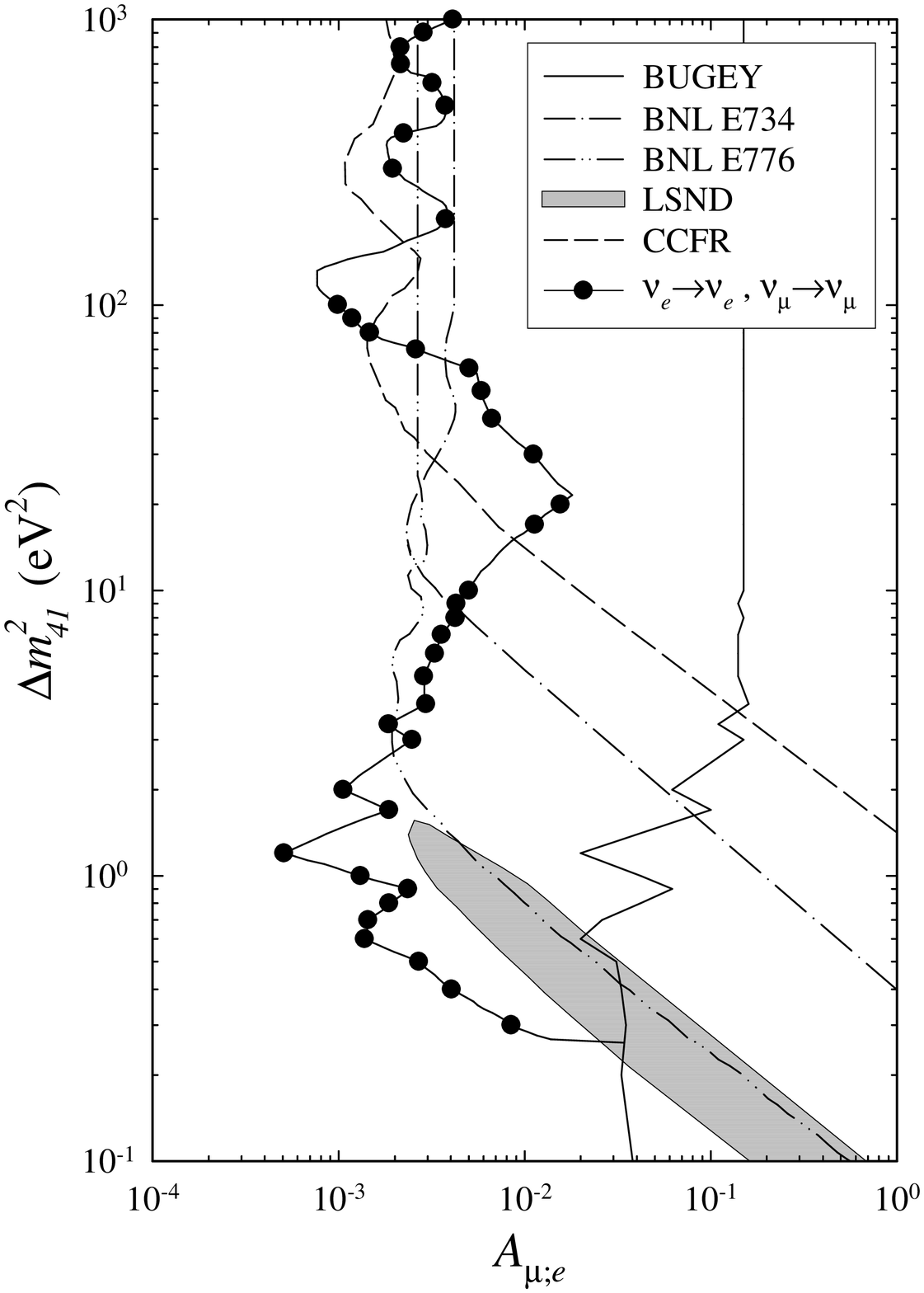,width=\linewidth}}
\\
Figure \ref{fig2}
\end{center}
\null \vspace{-1.5cm} \null
\refstepcounter{figure}
\label{fig2}
\end{figure}

In the case of spectra I and II, 
$\nu_\mu\to\nu_e$
transitions in SBL experiments are strongly suppressed.
Indeed, we have
\begin{equation}
A_{e;\mu}
\leq
4 \, |U_{ek}|^2 \, |U_{{\mu}k}|^2
\leq
4 \, a_e^0 \, a_\mu^0
\,.
\label{10}
\end{equation}
In Fig.~\ref{fig2} the upper bound
(\ref{10}) is compared with the latest LSND-allowed region (at 90\% CL).
Figure~\ref{fig2} shows that
the spectra of type I and II
(that include also the hierarchical spectrum)
are disfavored by the result of the LSND
experiment
(they are compatible with the results of the LSND experiment
only in the narrow region of $\Delta{m}^2_{41}$ around
$ 0.2 - 0.3 \, {\rm eV}^2 $,
where there is no information
on $B_{\mu;\mu}$).

On the other hand,
there is no incompatibility of
the spectra IIIA and IIIB with the results of the LSND experiment
and we conclude that these two spectra are favored
by the existing data.

We discuss now some consequences 
for future experiments that can be inferred from the schemes IIIA and IIIB.
Let us discuss first the possibilities
for the effective neutrino mass 
$m(^3{\rm H})$ measured in 
tritium $\beta$-decay experiments
and for the effective Majorana mass
$
|\langle{m}\rangle|
\equiv
\left| \sum_k U_{ek}^2 \, m_k \right|
$
measured in
neutrinoless double-$\beta$ decay experiments.

In scheme IIIA we have
\begin{equation}
m(^3{\rm H})
\simeq
m_4
\,,
\quad
|\langle{m}\rangle|
\leq m_4
\,,
\label{14}
\end{equation}
whereas
in scheme IIIB
\begin{equation}
\begin{array}{l} \displaystyle
m(^3{\rm H})
\leq
a_e^0 \, m_4 \ll m_4
\,,
\\ \displaystyle
|\langle{m}\rangle|
\leq
a_e^0 \, m_4 \ll m_4
\,.
\end{array}
\label{15}
\end{equation}
Therefore,
if the scheme A is realized in nature,
tritium $\beta$-decay experiments
experiments and
neutrinoless double-$\beta$ decay experiments
have a possibility to see the effects of the relatively
large neutrino mass $m_4 \simeq \sqrt{\Delta{m}^2_{{\rm LSND}}}$.

\begin{figure}[t!]
\begin{center}
\mbox{\epsfig{file=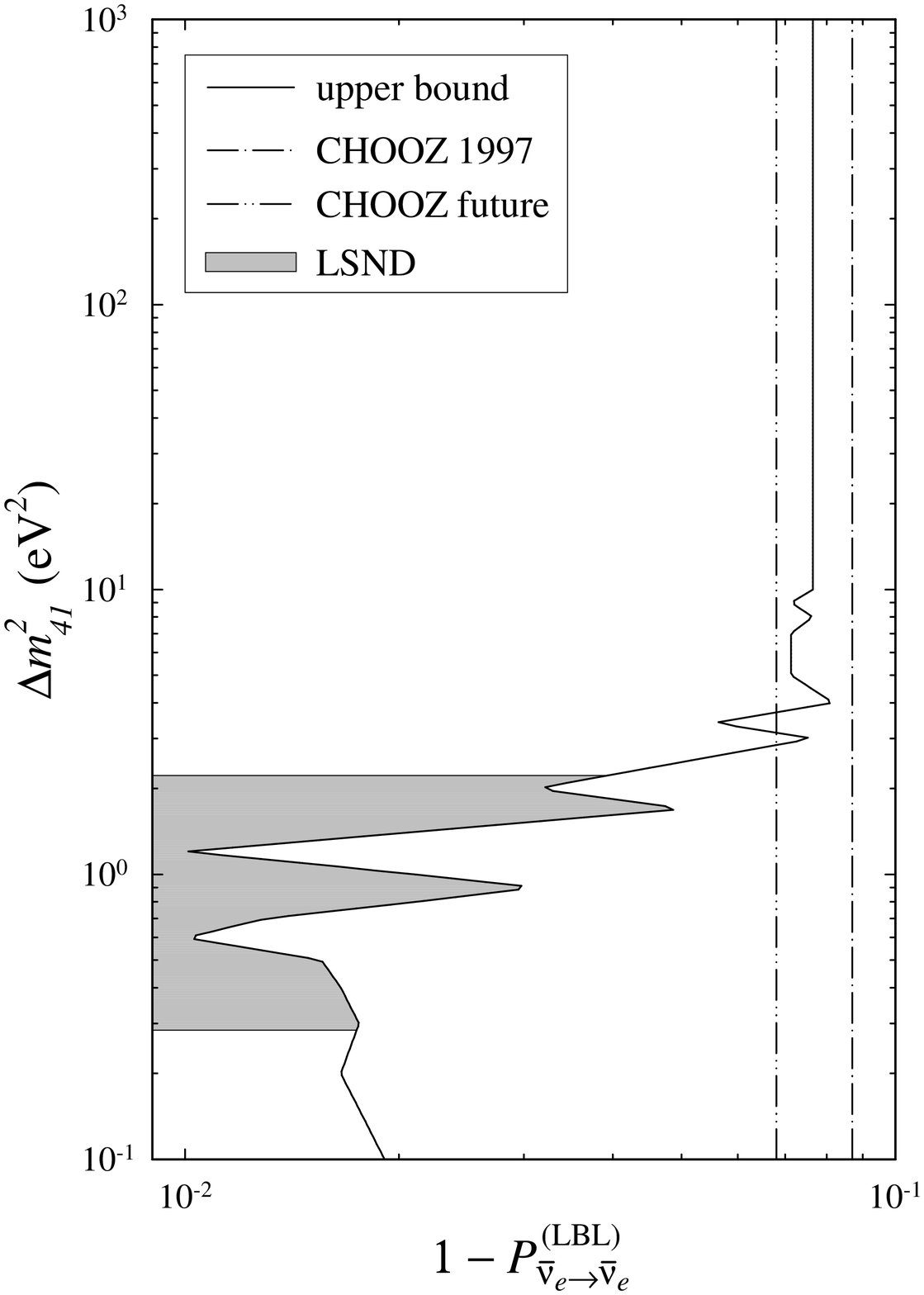,width=\linewidth}}
\\
Figure \ref{fig3}
\end{center}
\null \vspace{-1.5cm} \null
\refstepcounter{figure}
\label{fig3}
\end{figure}

Let us consider now neutrino transitions in long-baseline (LBL) experiments.
In the scheme IIIA the LBL transition probabilities are given by~\cite{BGG97a}
\begin{equation}
\begin{array}{rcl} \displaystyle
P_{\nu_\alpha\to\nu_\beta}^{{\rm LBL}}
&\displaystyle=& \displaystyle
\left| \sum_{k=1,2} U_{{\alpha}k}^* e^{-i\frac{\Delta{m}^2_{k1}L}{2E}} U_{{\beta}k} \right|^2
\\ \displaystyle
&& \displaystyle
+
\left| \sum_{j=3,4} U_{{\alpha}j}^* U_{{\beta}j} \right|^2
\,.
\end{array}
\label{16}
\end{equation}
The transition probabilities in the scheme IIIB can be obtained from
(\ref{16})
with the change $1,2\leftrightarrows3,4$.
The inequalities 
(\ref{08}) and (\ref{09})
imply strong constraints
on the probabilities of
$\bar\nu_e\to\bar\nu_e$
and
$\nu_\mu\to\nu_e$
transitions in LBL experiments~\cite{BGG97a}.
Indeed, for the probability of $\bar\nu_e\to\bar\nu_e$ transitions
we have
\begin{equation}
P_{\bar\nu_e\to\bar\nu_e}^{{\rm LBL}}
\geq
\left( \sum_{j=3,4} |U_{ej}|^2 \right)^2
\geq
(1-a_e^0)^2
\label{17}
\end{equation}
in scheme IIIA and
\begin{equation}
P_{\bar\nu_e\to\bar\nu_e}^{{\rm LBL}}
\geq
\left( \sum_{k=1,2} |U_{ej}|^2 \right)^2
\geq
(1-a_e^0)^2
\label{18}
\end{equation}
in scheme IIIB.
Hence,
in both schemes 
$P_{\bar\nu_e\to\bar\nu_e}^{{\rm LBL}}$
is close to one and the LBL probability of transitions of $\bar\nu_e$
into any other state is small.
Indeed,
in both schemes we have
\begin{equation}
1 - P_{\bar\nu_e\to\bar\nu_e}^{{\rm LBL}}
\leq
a_e^0 \, (2-a_e^0)
\,.
\label{19}
\end{equation}
This limit is shown by the solid line in Fig.~\ref{fig3}.
The upper bound for the transition probability
$ 1 - P_{\bar\nu_e\to\bar\nu_e}^{{\rm LBL}} $
obtained in the CHOOZ experiment~\cite{CHOOZ}
(dash-dotted line)
and the final sensitivity of the CHOOZ experiment
(dash-dot-dotted line) are also shown.
It can be seen that for
$ \Delta{m}^2_{41} \lesssim 1 \, {\rm eV}^2 $
the upper bound (\ref{19}) for
$ 1 - P_{\bar\nu_e\to\bar\nu_e}^{{\rm LBL}} $
is much smaller than
the upper bound reached in the CHOOZ experiment
and than the final sensitivity of the experiment.

\section{Three massive neutrinos}
\label{sec3}

If the results of the LSND experiment
will not be confirmed by future experiments,
the most plausible scheme is the one with mixing of three massive
neutrinos and a mass hierarchy
\cite{three,BGKM,BBGK96}:
\begin{equation}
m_1 \ll m_2 \ll m_3
\,.
\label{20}
\end{equation}
The effective Majorana mass
that characterize the matrix element of neutrinoless double-$\beta$
decay
is given in this case by~\cite{BGKM}
\begin{equation}
|\langle{m}\rangle|
\simeq
|U_{e3}|^2 \, \sqrt{\Delta{m}^2_{31}}
\,.
\label{21}
\end{equation}

\begin{figure}[t!]
\begin{center}
\mbox{\epsfig{file=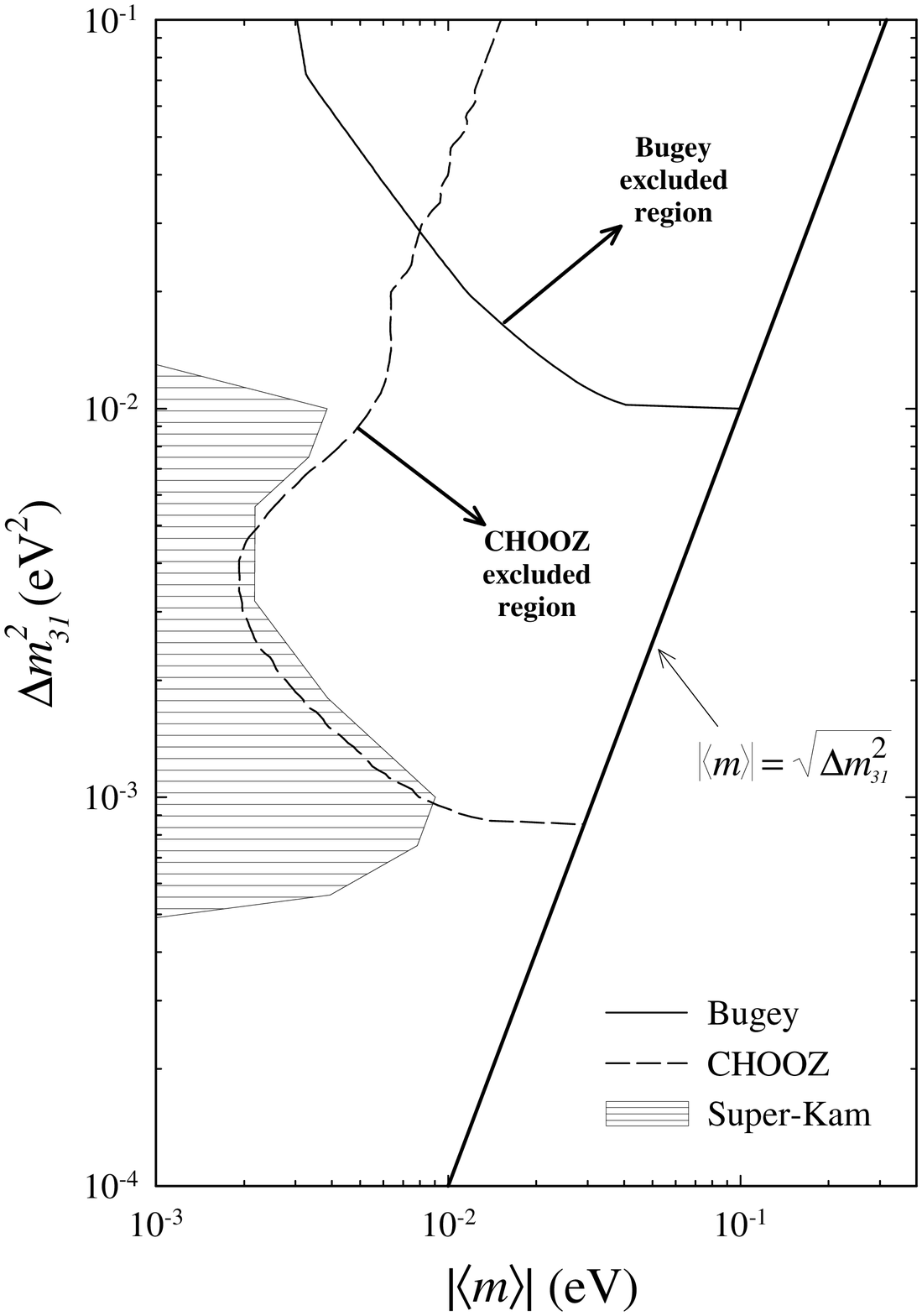,width=\linewidth}}
\\
Figure \ref{fig4}
\end{center}
\null \vspace{-1.5cm} \null
\refstepcounter{figure}
\label{fig4}
\end{figure}

The results of reactor neutrino experiments
imply the upper bound
$ |U_{e3}|^2 \leq a_e^0 $,
with $a_e^0$ given in Eq.(\ref{06}).
Therefore the effective Majorana mass
is bounded by
\begin{equation}
|\langle{m}\rangle|
\lesssim
a_e^0 \, \sqrt{\Delta{m}^2_{31}}
\,.
\label{22}
\end{equation}
The value of this upper bound as a function $\Delta{m}^2_{31}$
obtained from 90\% CL
exclusion plots of the Bugey~\cite{Bugey95} and CHOOZ~\cite{CHOOZ}
experiments
is presented in Fig.\ref{fig4}
(the solid and dashed line, respectively).
The region on the right of the thick straight solid line
is forbidden by the unitarity bound
$
|\langle{m}\rangle|
\leq
\sqrt{\Delta{m}^2_{31}}
$.

Also the results of
the Super-Kamiokande atmospheric neutrino experiment
imply an upper bound for
$|U_{e3}|^2$.
The shadowed region in Fig.\ref{fig4}
shows the
region allowed by Super-Kamiokande results at 90\% CL
that we have obtained
using the results of three-neutrino analysis performed by Yasuda~\cite{Yasuda}.

Figure \ref{fig4} shows that the results of the
Super-Kamiokande and CHOOZ experiments
imply that
$
|\langle{m}\rangle|
\lesssim
10^{-2} \, {\rm eV}
$.

The observation of neutrinoless double-$\beta$
decay with a probability
that corresponds to a value of
$|\langle{m}\rangle|$
significantly larger than
$10^{-2} \, {\rm eV}$
would mean that
the masses of three neutrinos do not have a hierarchical pattern
and/or exotic mechanisms (right-handed currents, supersymmetry
with violation of R-parity, \ldots, see~\cite{exotic})
are responsible for the process.

Let us notice that from the results of the Heidelberg-Moscow
$^{76}$Ge experiment~\cite{Heidelberg-Moscow}
it follows that
$
|\langle{m}\rangle|
\lesssim
0.5 - 1.5 \, {\rm eV}
$.
The next generation of neutrinoless double-$\beta$ experiments
will reach
$
|\langle{m}\rangle|
\simeq
10^{-1} \, {\rm eV}
$~\cite{next0bb}.
Possibilities to reach
$
|\langle{m}\rangle|
\simeq
10^{-2} \, {\rm eV}
$
are under discussion~\cite{next0bb}.

\section{Conclusions}
\label{sec4}

In conclusion,
the neutrino mass spectrum and the structure of the neutrino
mixing matrix depend crucially 
on the confirmation of the results of the LSND experiment.
If these results will be confirmed
we need (at least) four massive neutrinos
with a mass spectrum of type IIIA or IIIB
(see Fig.~\ref{fig1}).
If the results of the LSND experiment will not be confirmed,
the most plausible scenario is the one with
three massive neutrinos and a mass hierarchy.
The investigation of the nature of massive neutrinos (Dirac or Majorana?)
will require in this case to reach a sensitivity of
$ |\langle{m}\rangle| \lesssim 10^{-2} \, {\rm eV} $
in searching for
neutrinoless double-$\beta$ decay.

\end{document}